\documentclass[roman,12pt]{WileyMSP-template}

\def\scrystal{S1} 
\def\sxray{S2} 
\def\sesr{S3} 
\def\smag{S4} 
\def\sac{S5} 
\def\sph{S6} 
\def\sall{S7} 
\def\sFD{S8} 
\def\sMTHz{S9} 
\def\scalc{S10} 
\def\Sxm{S11} 
\def\SdisM{S12} 
\def\SdosM{S13} 
\def\sMM{S14} 
\def\sHS{S15} 
\usepackage{textcomp}
\usepackage{gensymb}
\usepackage{fancyhdr}
\usepackage{caption}
\usepackage{graphicx}
\usepackage{xcolor}
\usepackage{lmodern,bm}    
\usepackage{microtype}
\usepackage{pdfpages}

\usepackage[superscript,biblabel]{cite}
\usepackage{hyperref}
\definecolor{red}{rgb}{0,0,0}
\newenvironment{revision}{\par\color{red}}{\par}
\makeatletter \renewcommand{\@citess}[1]{\textsuperscript{\,[#1]}} \makeatother

\begin{document}

\pagestyle{fancy}
\fancyfoot[C]{\thepage}
\setlength{\emergencystretch}{1em}

\rhead{\includegraphics[width=2.5cm]{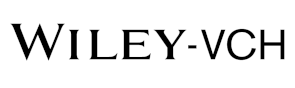}}

\title{Multi-Center Magnon Excitations Open the Entire Brillouin Zone to Terahertz Magnetometry of Quantum Magnets}
\maketitle

\medskip

\author{Tobias Biesner$^{\dagger}$,}
\author{Seulki Roh$^{\dagger}$,}
\author{Aleksandar Razpopov,}
\author{Jannis Willwater,} 
\author{Stefan Süllow,} 
\author{Ying Li,} 
\author{Katharina M. Zoch,} 
\author{Marisa Medarde,} 
\author{Jürgen Nuss,} 
\author{Denis Gorbunov,} 
\author{Yurii Skourski,}
\author{Andrej Pustogow,}
\author{Stuart E. Brown,}
\author{Cornelius Krellner,}
\author{Roser Valent\'{\i}$^*$,}
\author{Pascal Puphal$^*$,}
\author{Martin Dressel$^*$}

\dedication{$^\dagger$These authors contributed equally. $^*$Corresponding author}

\begin{affiliations}
T. Biesner, Dr. S. Roh, Ass. Prof. A. Pustogow, Prof. M. Dressel\\
1.~Physikalisches Institut, Universit\"at Stuttgart, Stuttgart, Germany\\
Email Address: dressel@pi1.physik.uni-stuttgart.de (M.D.)\\
\vspace{\baselineskip}
A. Razpopov, Dr. Y. Li, Prof. R. Valent\'{\i}\\
Institut für Theoretische Physik, Goethe-Universit\"at Frankfurt, Frankfurt am Main, Germany\\
Email Address: valenti@itp.uni-frankfurt.de (R.V.)\\
\vspace{\baselineskip}
J. Willwater, apl. Prof. S. Süllow\\
Institut f\"ur Physik der Kondensierten Materie, Technische Universit\"at Braunschweig, Braunschweig, Germany\\
\vspace{\baselineskip}
Dr. Y. Li\\
Department of Applied Physics and MOE Key Laboratory for Nonequilibrium Synthesis and Modulation of Condensed Matter, School of Physics, Xi’an Jiaotong University, Xi’an, China\\
\vspace{\baselineskip}
K. M. Zoch, Prof. C. Krellner, Dr. P. Puphal\\
Physikalisches Institut, Goethe-Universit\"at Frankfurt, Frankfurt am Main, Germany\\
\vspace{\baselineskip}
Dr. M. Medarde, Dr. P. Puphal\\
Laboratory for Multiscale Materials Experiments, Paul Scherrer Institute, 5232 Villigen PSI, Switzerland\\
\vspace{\baselineskip}
Dr. J. Nuss, Dr. P. Puphal\\
Max Planck Institute for Solid State Research, Stuttgart, Germany\\
Email Address: p.puphal@fkf.mpg.de (P.P.)\\
\vspace{\baselineskip}
Dr. D. Gorbunov, Dr. Y. Skourski\\
Hochfeld-Magnetlabor Dresden (HLD-EMFL), Helmholtz-Zentrum Dresden-Rossendorf, 01328 Dresden, Germany\\
\vspace{\baselineskip}
Ass. Prof. A. Pustogow, Prof. S. E. Brown\\
Department of Physics and Astronomy, UCLA, Los Angeles, USA\\
\vspace{\baselineskip}
Ass. Prof. A. Pustogow\\
Institute of Solid State Physics, TU Wien, 1040 Vienna, Austria
\end{affiliations}
\medskip
\clearpage
\fontsize{12bp}{18bp}\selectfont

\begin{abstract}
Due to the small photon momentum, optical spectroscopy commonly probes magnetic excitations only at the center of the Brillouin zone; however, there are ways to override this restriction.
In the case of the distorted kagome quantum magnet Y-kapellasite, Y$_3$Cu$_9$(OH)$_{19}$Cl$_8$, under scrutiny here, the \textcolor{red}{magnon} density of states can be accessed over the entire Brillouin zone through \textcolor{red}{three-center} magnon excitations. This mechanism is aided by the three different magnetic sublattices and strong short-range correlations in the distorted kagome lattice. The results of THz time-domain experiments agree remarkably well with linear spin-wave theory. {\color{red}Relaxing the conventional zone-center constraint of photons gives a new aspect to probe magnetism in matter.}

\end{abstract}

\section{Introduction}
\fontsize{12bp}{18bp}\selectfont
Conservations of momentum and energy are fundamental principles in physics that have to be obeyed also by optical excitations in solids. In an isolated system, for example, direct electronic transitions induced by light follow both $\Delta E = E_{ph}$ and  $\Delta \bm{q} =\bm{q}_{ph}$, where $\Delta E$ and $\Delta \bm{q}$ are the energy and momentum differences between the initial and the excited states, respectively, while $E_{ph}$ and $\bm{q}_{ph}$ are energy and momentum of the photon.\cite{dressel2002,Tanner2019} By the same token, these two principles dictate magnetic excitations as well. Here, a magnon with momentum $\bm{q}$ is created / annihilated by absorbing / emitting a photon. Within the energy conservation limit, the small photon momentum confines possible excitations close to the center of the Brillouin zone (BZ), i.e. $\bm{q}\approx0$, due to its sharp energy-momentum dispersion ($\omega = q_{ph}c_0/n$, $\omega$: angular frequency, $c_0$: speed of light in vacuum, $n$: refractive index of medium). Thus, optical accesses, for instance in Raman and THz spectroscopy, to the magnetic transitions are focused on the zone-center magnons.\textcolor{red}{\cite{Fishman2018,Wulferding2019,Zhang18,Little17,Chauhan2020,Laurita2015}} On the other hand, neutron scattering covers a broad range in the Brillouin zone due to large momenta of neutrons, promoting this technique as ideal spectroscopic tool to map spin dispersion. In reality, however, the picture is tainted by the low energy and wavevector resolution of neutron scattering experiments and restriction to large-scale facilities. Further constraints arise from the required large sample mass and the preference to avoid hydrogen and other isotopes with strong neutron absorption. \textcolor{red}{In turn, optical techniques could provide easy access to the magnetism albeit the $\bm{q}\approx0$ constraint. Attempts to overcome this constraint are subject of current research and first promising results have been reported.\cite{Nguyen2017,Wu2018,Hortensius2021}}

\begin{revision}
In THz spectroscopy, with the aid of ultrafast lasers and advanced detection schemes, dynamical processes in the (sub-) picosecond range become accessible.\cite{Walowski16,Eschenlohr17} This progress allows not only performing time-resolved studies but enables a control of magnetization dynamics.\cite{Nemec2018,Kampfrath13,Kimel05} Magnetic properties can be probed, for instance, through the magneto-optical Faraday or Kerr effects.\cite{Kampfrath11,Vicario13} Other examples include the coherent generation of magnons.\cite{Hortensius2021} Control over THz-driven spin precession \cite{Hansteen2005, Nakajima2010,Yamaguchi2010,Yamaguchi2013} opens the field of THz magnetometry, i.e. the fully optical extraction of magnetic properties.\cite{Zhang2020} \end{revision}

\begin{figure*}[t!]
\centering
\includegraphics[width=1.05\linewidth]{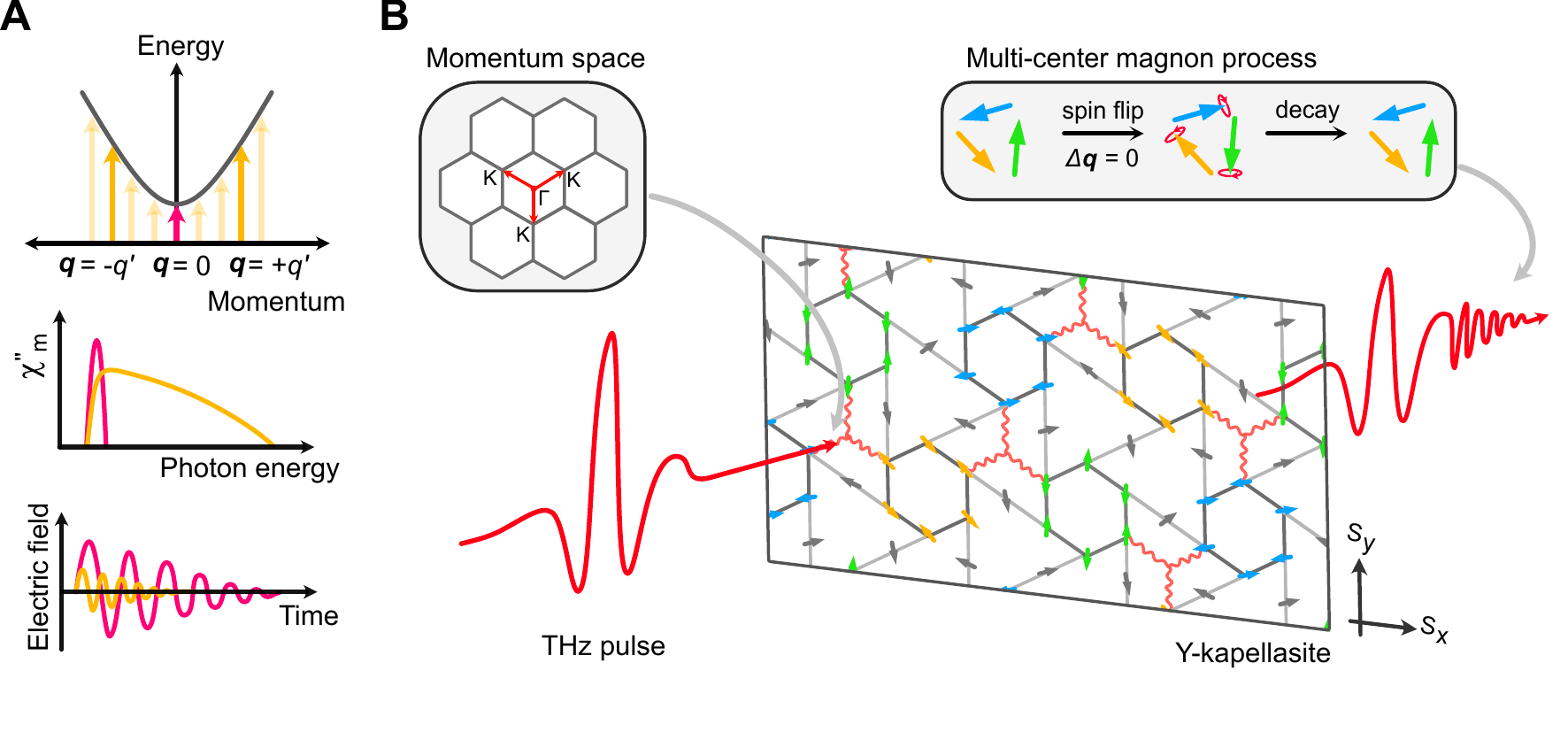}
\caption{\label{Fig0}\textcolor{red}{Three-center} magnon process in Y-kapellasite. a) Comparison between the one-magnon (red) and \textcolor{red}{two-center magnon} absorption (yellow). The top panel describes the excitations in the spin dispersion curve. While the one-magnon excitation only takes place near the zone center ($\bm{q} \approx 0$), the two-center magnon excitations can extend over the entire Brillouin zone ($\Delta \bm{q} \approx0$ excitation). Here, the summed momenta of the participating magnons need to vanish, $0\approx\sum_{n} \bm{q}_n$, to ensure momentum conservation. The middle and bottom panel show the corresponding response in energy / frequency-dependent magnetic susceptibility $\chi^{''}_m$ and electric field as a function of time, respectively. A one-magnon excitation shows a sharp peak in $\chi^{''}_m$, due to comparably long lifetimes (20-40 ps). \textcolor{red}{A multi-center magnon} process, however, is restricted to shorter time scales, \textcolor{red}{correlated with} a broad, continuum-like feature in $\chi^{''}_m$. b) Schematics of \textcolor{red}{three-center} magnon absorption in Y-kapellasite and calculated ground state, $\bm{Q}~=~(1/3,1/3)$. The simultaneous magnetic absorption occurs through three different magnetic sublattices (green, blue, and yellow hexagons in real space). The excited spin waves fall back to the initial state via a free induction decay resulting in oscillations of the outgoing THz pulse at extended time.}
\end{figure*}

Here, we demonstrate the potential of THz time-domain spectroscopy (THz-TDS) \cite{Neu2018}
in easing \textcolor{red}{its} zone-center restriction: \textcolor{red}{the spin (magnon) density of states (SDOS)} can be probed over the entire Brillouin zone via \textcolor{red}{multi-center magnon absorption} in materials with several magnetic sublattices. In this picture, magnon excitations take place simultaneously in different magnetic sublattices and positions in momentum space, $\bm{q}_n$.\textcolor{red}{\cite{Wright1969,Fleury1968,Loudon1968}} Yet, the summed momenta of the participating magnons \textcolor{red}{are} zero, $\sum_{n} \bm{q}_n \approx 0$, achieving momentum conservation. \textbf{Figure~\ref{Fig0}}a compares the conventional excitation picture, i.e.\ one-magnon process ($\bm{q}\approx0$), with the \textcolor{red}{multi-center magnon} process ($\Delta \bm{q}\approx0$). The distinction between both is prominent in the THz susceptibility $\chi^{''}_m$ and in the time-dependent electric field of the probing light. Here, the $\bm{q}\approx0$ mode (one-magnon) causes a homogeneous precession of spins,\textcolor{red}{\cite{Rezende2019,Nemec2018}} resulting in sharp spectral features in $\chi^{''}_m$ and relatively long oscillations of the electric field (20-40 ps) \textcolor{red}{\cite{Grishunin18 ,Lu17}} through a free induction decay mechanism.\textcolor{red}{\cite{Hansteen2005,Nakajima2010,Yamaguchi2010,Yamaguchi2013}} \textcolor{red}{The multi-center magnon absorption yields a broad feature \cite{Wright1969,Thorpe1970,Azuma05} implying a reduced lifetime (cf.\ red and yellow contributions in Figure~\ref{Fig0}a). It can be observed even above the magnetic long-range ordering temperature $T_N$ as a semblance of paramagnons (paramagnons are observed by various methods, see References~\cite{LeTacon2011,Qin2017,Kadlec11}).}

We have chosen the quantum magnet \cite{Balents2010,Mendels16,Kohama19,Norman16,Knolle2019} Y-kapellasite (Y$_3$Cu$_9$(OH)$_{19}$Cl$_8$) \cite{Puphal17} as our material platform. Y-kapellasite forms a distorted kagome lattice and has attracted attention recently due to its intriguing magnetic ground state ($T_N=2.2$ K), suggested in Figure~\ref{Fig0}b \cite{Hering2021}. Here, the magnetic supercell is comprised of three hexagons rotated by 120$^\circ$. \textcolor{red}{Due to the three possible arrangements, we define each hexagon as a magnetic sublattice and give it one of the colors: green, blue, or yellow.} Strong short-range magnetic correlations and persistent spin dynamics below the ordering temperature (down to $T=20$ mK) were reported previously.\cite{Puphal17,Barthelemy19} The THz transmission of Y-kapellasite was measured over a wide temperature and magnetic field range. At low temperatures ($T<30$ K), we find magnetically active, continuum-like excitations decaying through oscillations of the transient electric field during an extended time period of around 5 ps. By comparison to the linear spin-wave theory, we conclude that the THz $\chi^{''}_m$ encodes the SDOS through a \textcolor{red}{multi-center magnon absorption}, \textcolor{red}{three-center} magnon, augmented by the distinct magnetic ground state of Y-kapellasite, as described in Figure~\ref{Fig0}b. \textcolor{red}{We propose THz magnetometry via multi-center magnon absorption as a method to overcome the conventional zone-center restriction providing access to magnetism over the entire Brillouin zone.}
\clearpage
\section{Results}
\subsection{Magnetic properties of Y-kapellasite}
\begin{figure*}[t!]
\centering
\includegraphics[width=0.95\linewidth]{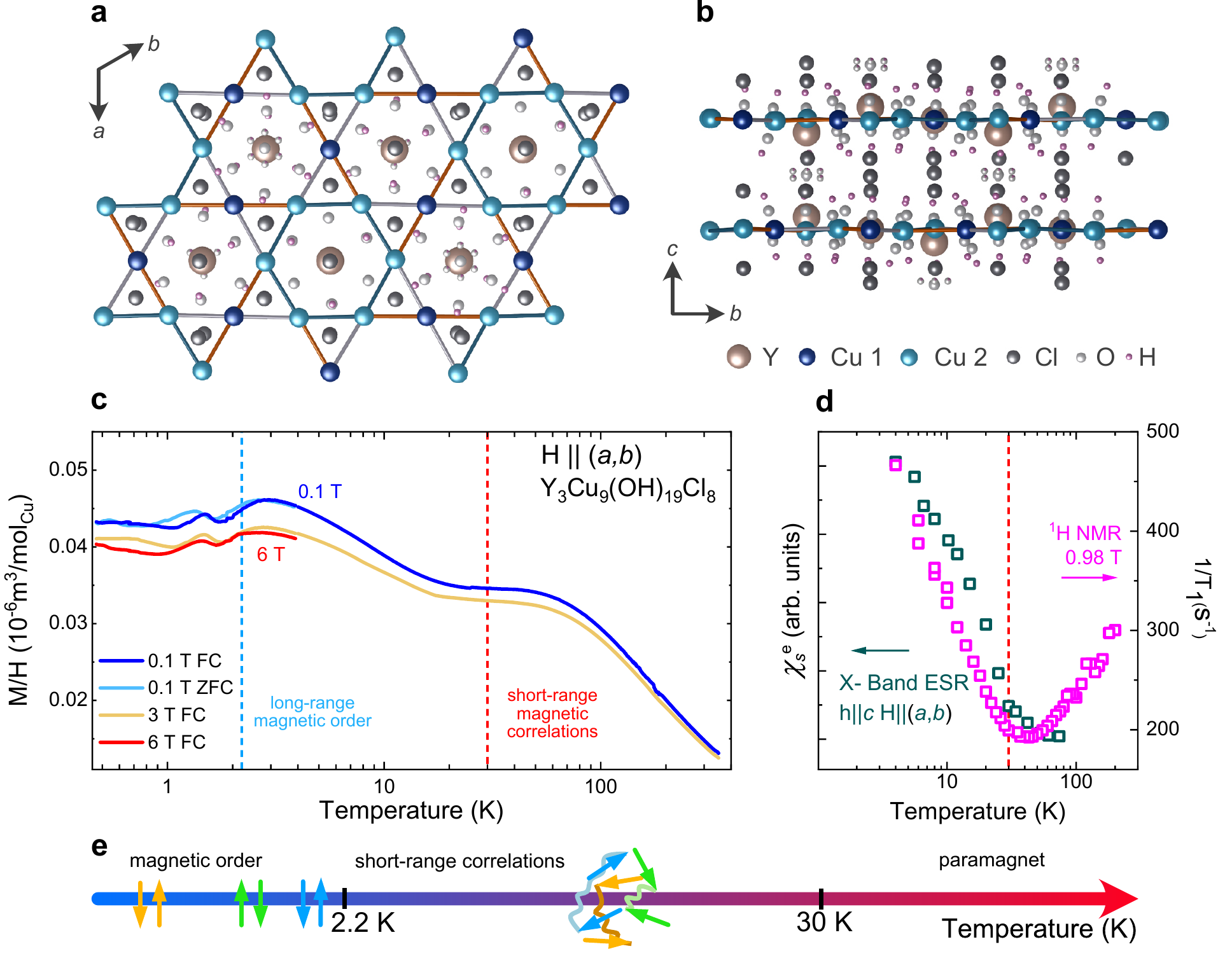}
\caption{\label{Fig1}Crystal structure and magnetic properties of Y-kapellasite. a) In-plane, \textcolor{red}{$(a,b)$-direction} and b) out-of-plane ($c$-axis) crystal structure. Different copper sites, inducing the distorted kagome bonds (blue, red, grey) are depicted in cyan and dark blue. c) Temperature-dependent dc magnetic susceptibility $M/H$ (FC - field cooled, ZFC - zero field cooled), under several in-plane magnetic fields. The red dashed line correspond to the onset of short-range magnetic correlations, the light blue dashed line indicates the onset of long-range magnetic order. d) Spin-lattice relaxation rate $1/T_1$ ($^1$H NMR, right axis) and electron spin susceptibility $\chi_s^e$ (ESR, left axis). Red dashed line: Onset of short-range magnetic correlations. e) Extracted temperature ranges, onset of short-range magnetic correlations ($30$~K) and long-range magnetic order (below $2.2$~K).}
\end{figure*}

In Y-kapellasite, a distorted kagome lattice is formed by two distinct Cu sites, as depicted in \textbf{Figure~\ref{Fig1}}a. Here, the magnetic superexchange is governed along three different Cu-O-Cu paths.\cite{Puphal17,Hering2021} The distorted kagome planes stack along the $c$-direction (Figure~\ref{Fig1}b). See supplement for further material information (\textbf{Figure~\scrystal~and~\sxray}). Let us first consider the static response of the spin system. The in-plane magnetic susceptibility ($\bm{H}\parallel (a,b)$) down to temperatures as low as $T=400$~mK is presented in Figure~\ref{Fig1}c.
Focusing on the $H=0.1$~T measurements, upon cooling, the magnetization first saturates in a plateau at intermediate temperatures ($T=50$~K) and continues to rise below $T=30$~K. At around $T=3$~K a maximum develops, followed by a smaller peak at lower temperature ($T=1.5$~K). While the maximum at $T=3$~K is consistent with the previously reported $T_N=2.2$ K,\cite{Puphal17} also observed in heat-capacity measurements, the additional smaller peak suggests a successive freezing of the magnetic texture at lower temperatures. This is in accordance with the recent $\mu$SR study on polycrystalline samples,\cite{Barthelemy19} reporting persistent spin dynamics down to the mK range. Furthermore, a weak hysteresis between field cooled (FC) and zero field cooled (ZFC) measurements is observed below $T=3$~K, whereas the cooling protocol does not affect the susceptibility noticeably at higher temperatures, indicating the contribution of uncompensated spins at low temperatures. Increasing the external magnetic field, the magnetization is slightly suppressed and the 3~K-peak shifts to lower temperatures, implying dominant antiferromagnetic interactions. This result suggests two different characteristic temperatures in Y-kapellasite, around $T=30$~K and $T_N=2.2$~K.

For additional information, the magnetic properties of Y-kapellasite are further investigated using nuclear magnetic resonance (NMR) and electron spin resonance (ESR). Starting with the $^1$H-NMR characterization (Figure~\ref{Fig1}d), the in-plane spin-lattice relaxation rate $1/T_1$ first decreases upon cooling and increases again near $T=30$~K. This crossover indicates the onset of short-range magnetic correlations. A similar temperature scale is found by ESR measurements as shown in Figure~\ref{Fig1}d, left axis (\textbf{Figure~\sesr}~for extended ESR spectra). \textcolor{red}{Simultaneously} with the onset of magnetic correlations at around $T=30$~K, the electron spin susceptibility $\chi_s^{e}$ starts to rise, advocating the close relation between the ESR absorption and the short-range magnetic interactions. Additional measurements, high-field magnetization and ac susceptibility, \textcolor{red}{reveal} similar temperature scales and are presented in the supplement (cf.\ \textbf{Figure~\smag~and~\sac}). The range between $T\approx30$~K (short-range magnetic correlations) and $T_N\approx2.2$~K offers experimental access for investigating the emergence of magnetism on the kagome lattice above $T_N$, as sketched in Figure~\ref{Fig1}e. The desired \textcolor{red}{multi-center magnon absorption} could survive for higher temperatures through these short-range magnetic correlations.

\subsection{THz time-domain spectra as a probe of spin dynamics}
\begin{figure*}[t!]
\centering
\includegraphics[width=1\linewidth]{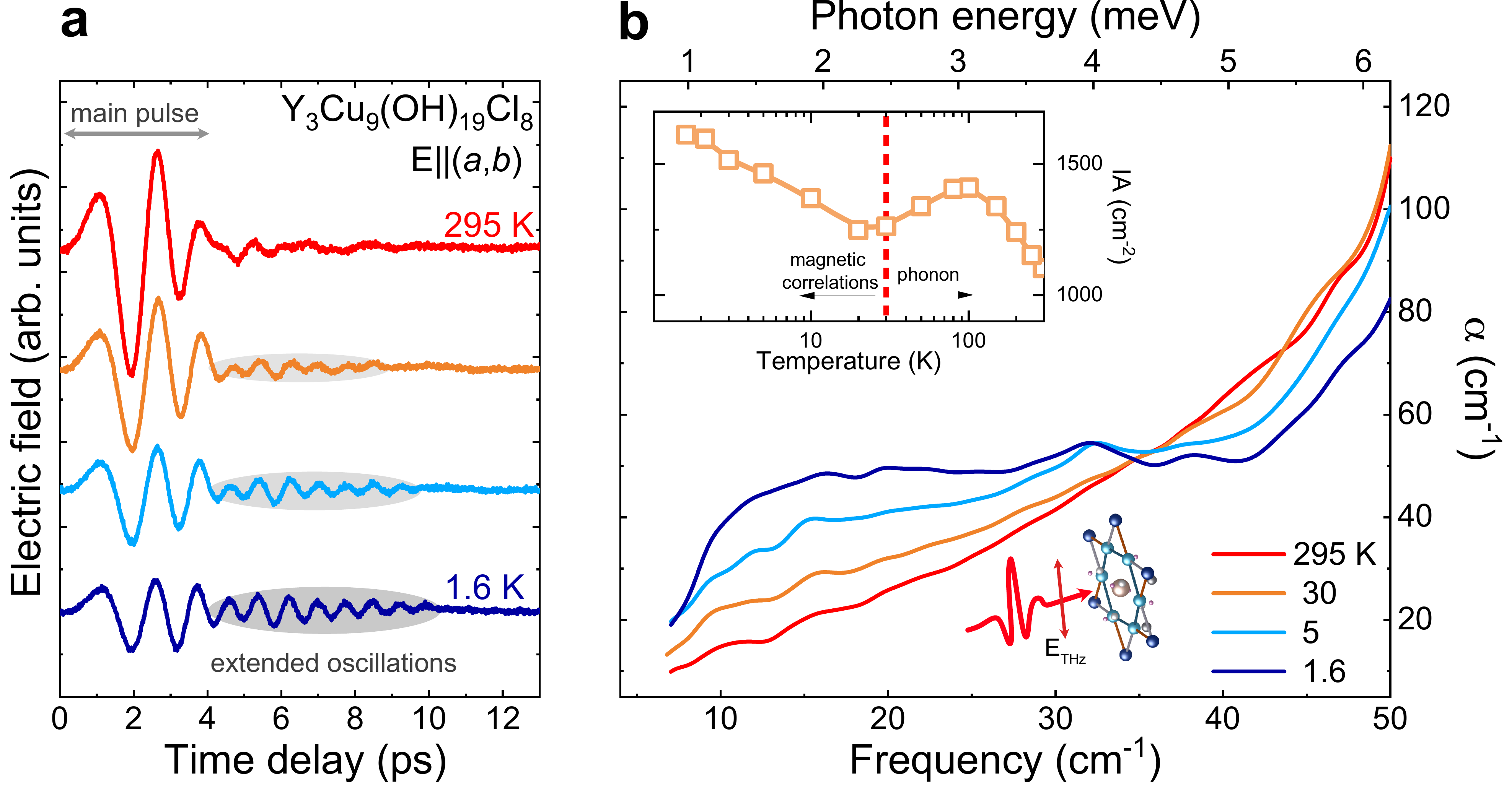}
\caption{\label{Fig2}Temperature-dependent THz spectra. a) THz electric field, transmitted through Y-kapellasite as a function of time delay and b) resulting absorption coefficient $\alpha$ for the in-plane direction, as depicted ($\bm{E}_{THz}$ $\parallel (a,b)$). At $T=295$~K, the main pulse (0 to 4 ps) contains most of the THz responses (phononic properties). Below the onset of short-range magnetic correlations, $T=30$~K, the THz electric field pronounces extended-time oscillations (4 to 10 ps, shaded area), while the main pulse loses its intensity. As a result the continuum-like absorption develops in $\alpha$, followed by two additional peak-like features at 32 and 37 cm$^{-1}$. Inset: Integrated absorption coefficient $\textit{IA}$ (up to $40$~cm$^{-1}$). The red dashed line represent the onset of short-range magnetic correlations, $T=30$~K. Above $T=30$~K, changes in $\textit{IA}$ are mostly caused by the lowest in-plane phonon mode. Below the onset of short-range magnetic correlations, the $\textit{IA}$ continuously increases.}
\end{figure*}

In \textbf{Figure~\ref{Fig2}} we plot the raw data of the THz-TDS \textcolor{red}{recorded} in the $ab$-plane.
At high temperatures, the in-plane time-domain signal of the transient electric field (Figure~\ref{Fig2}a)
consists of only the main pulse ranging from 0 to 4 ps (\textcolor{red}{see Methods}).
With cooling, the intensity of this pulse decreases,
but in addition a pronounced oscillating electric field develops over an extended time ranging from 4 to 10 ps.
\textcolor{red}{Note, for conventional one-magnon excitations, a similar time-domain signal was reported but on a significantly longer time scale (several tens of ps).\cite{Lu17,Grishunin18,Nakajima2010,Yamaguchi2010,Yamaguchi2013}} To gain more insight into the underlying physical processes,
we perform a Fourier transformation and calculate the frequency-dependent absorption coefficient \textcolor{red}{$\alpha$, plotted in Figure~\ref{Fig2}b}.
At room temperature, only the tail of the lowest in-plane phonon mode contributes notably to the THz absorption.
Upon cooling, the phonon contribution first increases slightly and then becomes weaker together with a suppression of the main pulse in the time-domain signal (see supplement, \textbf{Figure~\sph}~\textcolor{red}{for the results} of infrared spectroscopy and DFT calculations of the phonons).
As the temperature drops below $T=30$~K, we see the low-frequency absorption rising, \textcolor{red}{resulting in a broad continuum-like contribution much stronger than the phonon (in this frequency range).}
This corresponds to the enhancement of the electric field oscillations at extended times 
as illustrated in Figure~\ref{Fig2}a.
The feature increases strongly with further cooling and finally dominates the entire THz response down to the lowest temperature measured.  
Furthermore, for $T < 5$~K, we observe two weaker, but noticeable, peak-like contributions between $30-40$~cm$^{-1}$. Despite the increase of intensity, no qualitative difference can be found in measurements below $T_N$. From the comparison with our calculations, we can exclude a phononic origin of the low-energy features. Also electronic contributions are unlikely for the highly insulating Y-kapellasite (bandgap of 3.6 eV).\cite{Puphal17,Pustogow17}

These trends become even more obvious when looking at the integrated absorption $\textit{IA}$ (similar to the spectral weight, see method section), displayed in the inset of Figure~\ref{Fig2}b.
The crossover range in $\textit{IA}(T)$ matches well with the onset of short-range magnetic correlations at around $T=30$~K as observed in the NMR spin-lattice relaxation rate and ESR susceptibility.
This good agreement between the temperature scales strongly suggests
that the THz continuum-like absorption is caused by short-range magnetic correlations even above $T_N$ (see supplemental \textbf{Figure~\sall}\
for corresponding spectra covering the entire temperature range; \textbf{Figure~\sFD}~ for additional data in the frequency domain).

\subsection{Magneto-THz spectroscopy}
\begin{figure*}[t!]
\centering
\includegraphics[width=1\linewidth]{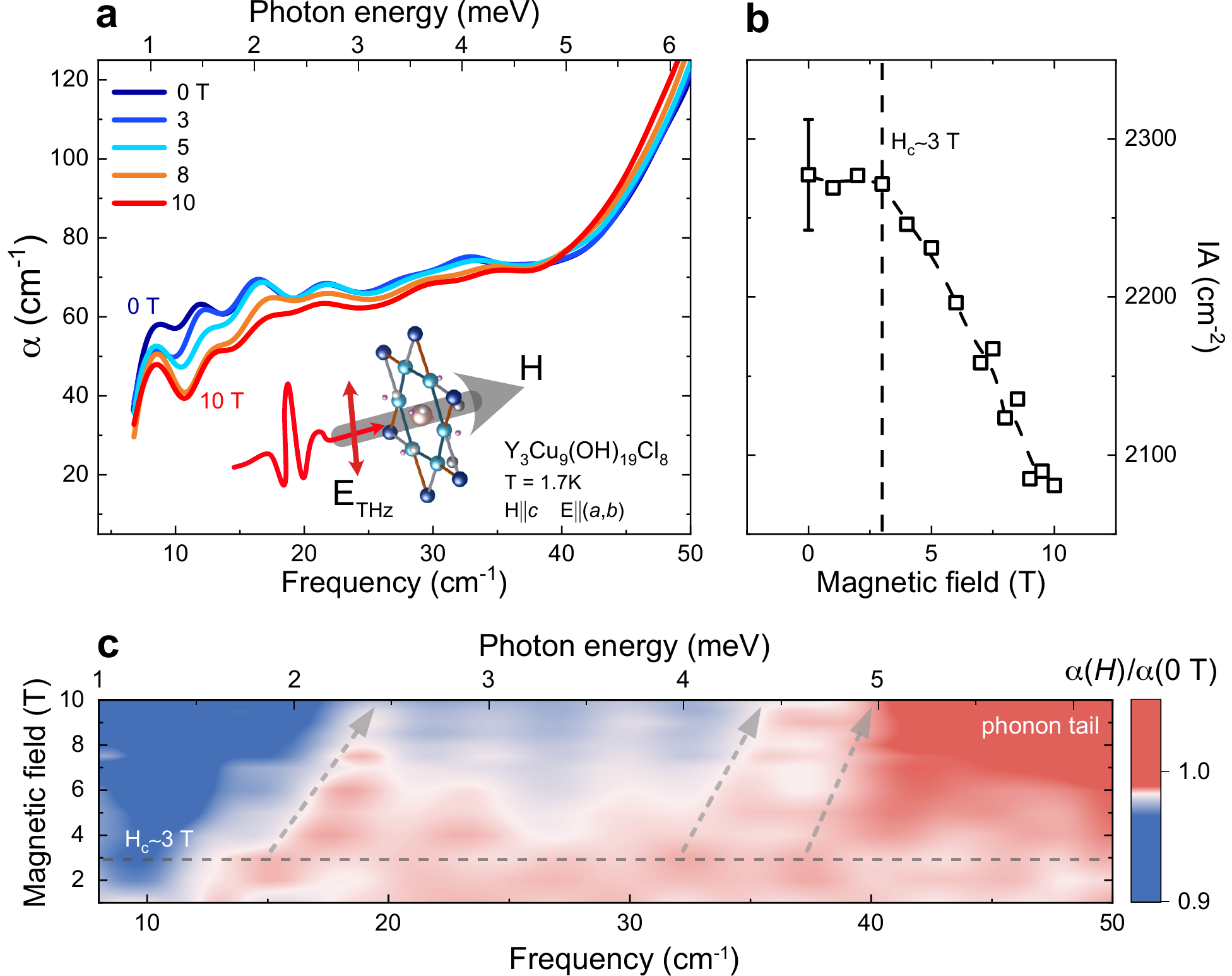}
\caption{\label{Fig3}Result of magneto-THz spectroscopy at 1.7~K. a) Absorption coefficient $\alpha$ under magnetic field in Faraday geometry, as depicted ($\bm{H}\parallel c$, $\bm{E}_{THz}\parallel (a,b)$). The THz continuum-like absorption decreases with increasing magnetic fields, confirming its magnetic origin. b) Integrated absorption coefficient $\textit{IA}$ (up to $40$~cm$^{-1}$), exposing a critical magnetic field of $H_c\sim3$~T. c) Contour plot of the relative absorption coefficient under magnetic field $\alpha(H)/\alpha(0~\mathrm{T})$, normalized to zero field. The grey arrows indicate the field evolution of the onset of the contiuum-like absorption and of the two peak-like features at 32 and 37 cm$^{-1}$.}
\end{figure*}

To clarify the magnetic origin of this continuum-like absorption, we carried out magneto-THz measurements at $T=1.7$~K while applying a static magnetic field up to $H=10$~T in Faraday geometry. With ramping up the magnetic field, the continuum-like feature loses its intensity beginning at the low-frequency end, as displayed in \textbf{Figure~\ref{Fig3}}a (\textbf{Figure~\sMTHz}~for extended data).
This can be nicely seen in Figure~\ref{Fig3}b where the $\textit{IA}$ is plotted as a function of $H$.
Exceeding a critical field of $H_c=3$~T, $\textit{IA}(H)$ decreases considerably;
at a field strength of $H=10$~T the intensity is reduced by almost 10~$\%$. This quantitative change seems reasonable since the magnetic energy corresponds to roughly 10~$\%$ of the exchange energy: $\mu_BH=0.1\,k_B\Theta_{CW}\approx0.1\,J$, where $\Theta_{CW}\approx-100$ K is the Curie-Weiss temperature and $J\approx13$ meV is the dominating exchange energy ($\mu_B$ is Bohr magneton, $k_B$ is Boltzmann constant).\cite{Puphal17,Hering2021} 
The phonon tail above 40~cm$^{-1}$ is affected as well: it weakly shifts towards lower energies.
In addition to these obvious changes in the spectrum,
we can identify some more subtle variations inside the continuum-like absorption.
Slight shifts of \textcolor{red}{the} $30-40$~cm$^{-1}$ peaks become clearer in the contour plot of the normalized absorption coefficient $\alpha(H)/\alpha(0~\mathrm{T})$, presented in Figure~\ref{Fig3}c.
Interestingly, the onset of the continuum-like absorption shows a stronger change under magnetic field, compared to the $30-40$~cm$^{-1}$ peaks, c.f.\ grey arrows in Figure~\ref{Fig3}c.
Note the reduced signal-to-noise ratio of a magneto-optical measurement leads to somewhat noisy features below 20~cm$^{-1}$; nevertheless, our magneto-optical THz results strongly support the magnetic origin of the continuum-like feature.

\subsection{\textcolor{red}{Multi-center magnon} excitations}
\begin{figure*}
\centering
\includegraphics[width=1\linewidth]{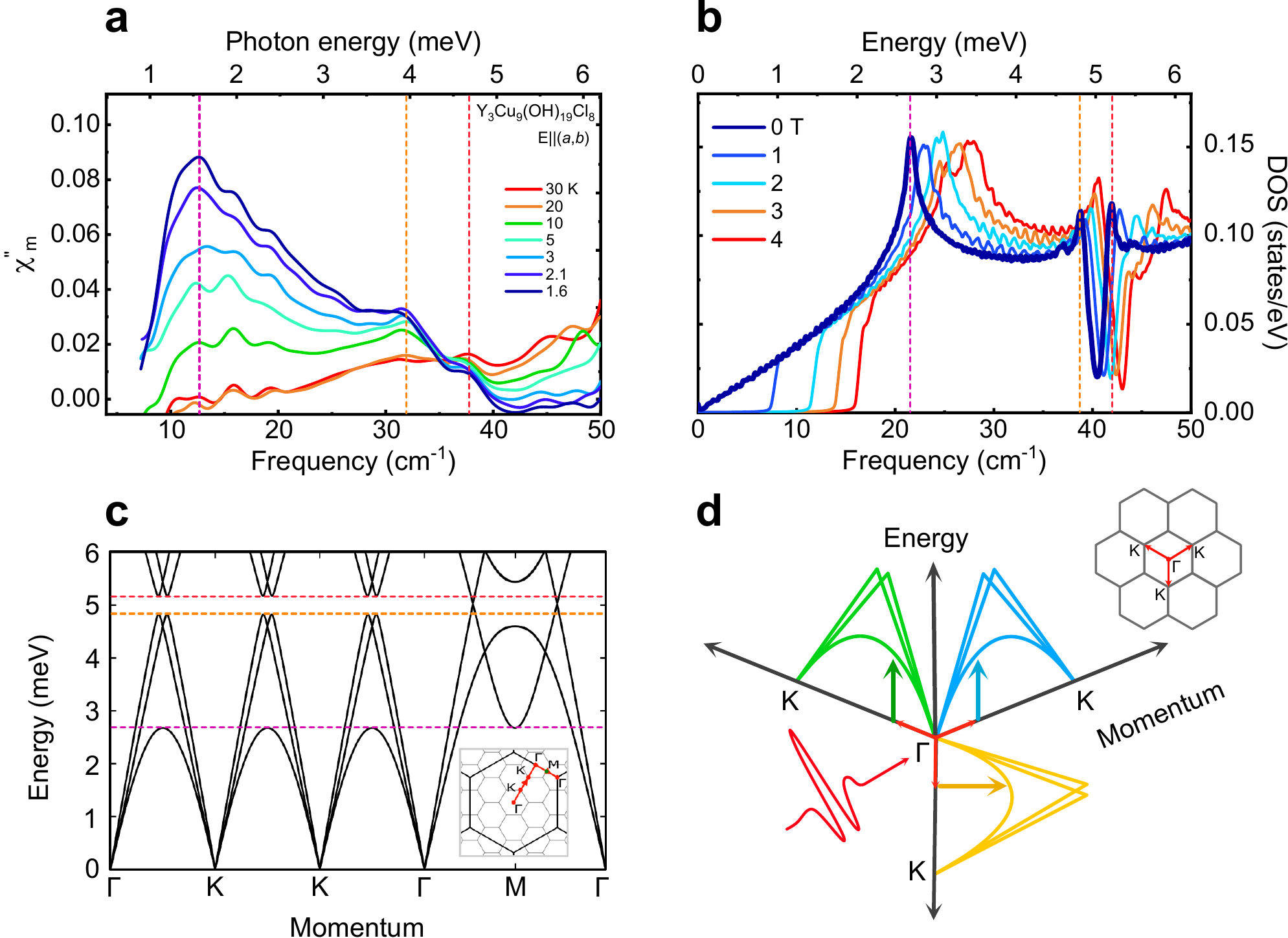}
\caption{\label{Fig4}\textcolor{red}{Three-center} magnon excitations and theoretical spin-wave dispersion of Y-kapellasite. a) The frequency-dependent imaginary part of the magnetic susceptibility $\chi^{''}_m$ obtained from THz-TDS exposes the natural spectral form of the \textcolor{red}{three-center} magnon excitations. Three distinctive features are marked by vertical dashed lines. b) Spin density of states (SDOS) obtained from linear spin-wave theory calculations \textcolor{red}{with and without} magnetic field. The SDOS shows three characteristic energies below 6 meV with a high density of states (vertical dashed lines). c) Spin-wave dispersion of Y-kapellasite. Horizontal dashed lines: Corresponding to the energies in (b). Inset: Calculated path in the extended Brillouin zone (black hexagon). d) The \textcolor{red}{three-center} magnon process in momentum space, i.e. three spin excitations in the different magnetic sublattices (green, blue, and yellow color code) with $0\approx\sum_{n} \bm{q}_n$.}
\end{figure*}

In a next step we have to separate the magnetic contributions from the dielectric ones in the THz absorption
in order to reveal their natural spectral form.
With the fair assumption that the dielectric properties do not change drastically at low temperatures, the magnetic susceptibility $\chi^{''}_{m}$ can be obtained from the THz spectra (Figure~\ref{Fig2}) by referencing to the 80~K-spectrum (see method section for further information). The results are shown in \textbf{Figure~\ref{Fig4}}a.
The extraction of $\chi^{''}_{m}$ unveils an asymmetric shape of the magnetic continuum-like feature
with a maximum at around 12~cm$^{-1}$ and a width of about 30~cm$^{-1}$ followed by two peaks at 32 and 37~cm$^{-1}$ at the lowest temperature, $T=1.6$~K.

To learn more about the origin of the magnetic THz response and the magnetic ground state, we calculated the spin-wave dispersion for Y-kapellasite with linear spin-wave theory by assuming a coplanar non-collinear $\bm{Q}= (1/3,1/3)$ magnetically ordered ground state as suggested by recent {\it ab-initio} DFT calculations.\cite{Hering2021} This ordered state comprises three different magnetic sublattices of hexagons (shown as green, blue, and yellow in Figure~\ref{Fig0}b). Within each hexagon, the spins are antiferromagnetically coupled and neighboring hexagons are coupled to each other via spins (shown in grey) whose directions are fully determined by the neighboring hexagons (slave spins).\cite{Hering2021}

Figure~\ref{Fig4}b~and~c contain the calculated spin density of states (SDOS) and the corresponding spin-wave dispersion in the relevant energy range, below $6$~meV (1 meV corresponds to 8.065~cm$^{-1}$). Several energies with a density of states, corresponding to characteristic energy scales of the THz $\chi^{''}_{m}$, are marked by dashed lines.
The spin-wave dispersion of Y-kapellasite is gapless and degenerated at the $\Gamma$-point (Goldstone mode);
possible one-magnon excitations are outside our THz spectral range (\textbf{Figure~\scalc}).
As introduced above, a one-magnon process only measures the response at the zone center;
in stark contrast to \textcolor{red}{multi-center magnon excitations} that can expand over the entire Brillouin zone.
For the latter case, due to momentum conservation, the total wavevector of the participating magnons must be zero, $\sum_{n} \bm{q}_n\approx0$, which is accomplished by the simultaneous magnon excitations in three \textcolor{red}{distinct} magnetic sublattices. Within this picture, the steady development of the continuum-like absorption without any additional features across \textcolor{red}{$T_N$} can be explained by \textcolor{red}{multi-center magnon absorptions}. The one-magnon excitations are outside our \textcolor{red}{accessible} spectral range and only the \textcolor{red}{multi-center magnons} are observed. As illustrated in Figure~\ref{Fig4}d, the spin-wave dispersion \textcolor{red}{beyond} the $\Gamma$-point gets accessible via such excitations. Indeed, towards the K-direction the degeneracy is lifted and the lowest band exhibits a parabolic shape (Figure~\ref{Fig4}c).
Around $2.7$~meV the band flattens and reaches a maximum. This shape of the dispersion yields a large number of available states in the low-energy range and generates the maximum in the SDOS at $2.7$~meV  which is related to the peak at $1.5$~meV (12~cm$^{-1}$) in $\chi^{''}_{m}$.
At higher energies, around $4.8$~meV, the density of states increases again with the next two higher-lying bands reaching their maxima. At $5$~meV the number of available states drops significantly before it peaks again at around $5.2$~meV (parabola minima of the next higher bands). Indeed, these states are related to the peak-like contributions between $32$ and $37$~cm$^{-1}$ ($4$ and $4.6$~meV) in $\chi^{''}_{m}$ (Figure~\ref{Fig4}a). There is a small offset of about $1$~meV between experimental results and calculations, probably due to a small mismatch in parameters between experiment and theory.

Overall, including the \textcolor{red}{multi-center magnon} picture, the SDOS from the linear spin-wave calculations show a remarkable agreement with the experimental result, $\chi^{''}_{m}$. \textcolor{red}{Furthermore, based on the SDOS, we calculated the expected $\chi^{''}_{m}$, validating the observed spectral shape (cf. \textbf{Figure~\Sxm}).} \textcolor{red}{The agreement with theory} even extends to the magnetic field dependence. Figure~\ref{Fig4}b displays the change of the SDOS with magnetic field. Under external magnetic field, the spin dispersion becomes gapped and the weight of the SDOS shifts up in energy (\textbf{Figure~\SdisM~and~\SdosM}).
In addition, the $2.7$~meV maximum reveals a strong shift under magnetic field, it is slightly different than the one obtained for the $4.8$~meV and $5.2$~meV peaks. We note that the calculations overestimate 
the magnetic field scale compared to that of the experiment, perhaps due to demagnetization effects. Still, these trends under external magnetic field are corroborating the result of our magneto-THz spectroscopy, displayed in Figure~\ref{Fig3}.
\clearpage
\section{Discussion}
Our THz study \textcolor{red}{allows} to directly probe the spin density of states via optical spectroscopy;
by utilizing \textcolor{red}{multi-center magnon absorption} we gain access to the SDOS over the entire Brillouin zone, i.e. \textcolor{red}{we actually perform} THz magnetometry. In Y-kapellasite, strong short-range magnetic correlations lead to spin fluctuations over a wide temperature range.
The paramagnonic behavior is aided by these fluctuations explaining the development of the THz continuum-like absorption together with the onset of short-range magnetic correlations. The lattice distortion, on the other hand, lowers the \textcolor{red}{spatial} symmetry and generates three magnetic sublattices.
This distortion is decisive for \textcolor{red}{multi-center magnon excitations} to occur. \textcolor{red}{In particular, the magnetic superstructure favors simultaneous $3N$ magnon excitations where each of the participating magnon excitations occurs, respectively, at one of the three magnetic sublattices, i.e.\ \textcolor{red}{three-center} magnon, while in total the net momentum is conserved ($\Delta \bm{q}= \bm{q}_{ph}\approx0$), as illustrated in Figure~\ref{Fig0}b (see supplement for further discussion, \textbf{Figure~\sMM}). We suggest that the role of distortion could be further verified in other kagome lattices. \cite{Hering2021,Watanabe2016,Ishikawa2019,Boldrin2018} But for the case of Y-kapellasite, the distortion seems to be crucial for the multi-center magnon absorption given that Herbertsmithite, which is considered as a perfect kagome structure,\cite{Norman16} does not reveal any signature of \textcolor{red}{multi-center magnons} (\textbf{Figure~\sHS}).\cite{Pilon13,Potter13}}

Such a symmetry breaking can also be phonon-assisted.\cite{Windt01,grueninger20} We note that similar suggestions were made for 
the parent compound of cuprates with an underlying square lattice. In those cases, the observed features were associated with the coupling between magnons and phonons leading to enhanced excitation frequencies in the mid-infrared range.\cite{Lorenzana1995,Perkins1993} The magneto-elastic coupling may be represented in the temperature dependence of the phonon modes. In the present case, we observe unusual redshifts and anomalies of the infrared-active phonons accompanied by the development of short-range magnetic correlations (Figure~\sph).

Overall, the comparison of our experimental results with linear spin-wave calculations suggests that for a proper description a \textcolor{red}{multi-center magnon} picture is required: we have to go
beyond a simple magnetic mode in order to explain the THz excitations.
In fact, the \textcolor{red}{multi-center magnon absorption} in Y-kapellasite resembles the conventional \textcolor{red}{two-center magnon} absorption observed in classical antiferromagnets: FeF$_2$, MnF$_2$, CoF$_2$, and NiF$_2$.\textcolor{red}{\cite{Halley1965, Tanabe1965, Allen1966, Loudon1968, Thorpe1969, Wright1969, Thorpe1970}}
Utilizing Dexter's theory of cooperative optical absorption,\cite{Dexter1962}
earlier reports proposed several mechanisms,\cite{Halley1965,Tanabe1965,Halley1967} where the removal of centrosymmetric points is supposed to be particularly necessary
for the non-vanishing electric dipole moment.\cite{Loudon1968}

The former studies confined the \textcolor{red}{two-center magnon} absorption as electric dipole active phenomena; but there is no reason for this limitation. Recently, a direct coupling between the magnetic state and light was proposed; the detection takes place over the free induction decay.\textcolor{red}{\cite{Hansteen2005,Nakajima2010,Yamaguchi2010,Yamaguchi2013}} As well, accounting for the two magnetic sublattices in conventional antiferromagnets, traditionally only a two-center absorption was considered.
Moreover, the optical selection rule for the \textcolor{red}{three-center} magnon excitation might differ from the conventional one-magnon picture ($\Delta S = \pm1$, the spin difference between initial and excited state). For example, the two-center magnon excitation was discussed before with an altered optical selection rule.\cite{Lohr1972,Tanabe2005} All these now turn into limiting factors when a more detailed and generalized microscopic picture
has to be established for the \textcolor{red}{multi-center magnon absorption}. Together with a
refined understanding, our approach might be applicable in a large pool of materials.

\begin{revision}
We further discuss the relation of the THz multi-center magnon absorption to similar multimagnon processes observed in other quantum magnets of low symmetry. For quantum magnets, often a virtual process via the magnon decay \cite{Zhitomirsky13} is discussed, which has been observed in Raman/ neutron scattering.\cite{Thompson2017,Hong2017,Wulferding2020,Sahasrabudhe20} Note that multimagnon scattering (photon/ neutron scatters by creating multiple magnons) and THz multi-center magnon absorption (multi-center magnon: one photon gets absorbed, simultaneously creating multiple magnons in different magnetic sublattices) need to be discerned with distinct optical selection rules.\cite{Loudon1968} Moreover, the higher-order magnon contributions beyond linear spin-wave theory are as well discussed in THz absorption. \cite{Pan2014} However, these are different from the multi-center magnon absorption, subject of our study. Regarding higher-order magnons, mixing of one-magnon branches with multimagnon states can occur as a result of the low symmetry of the spin interactions as in $\alpha$-RuCl$_3$. \cite{Winter2017,Wang17} A further example can be found in the case of Yb$_2$Ti$_2$O$_7$, showing a field-induced decay of the one-magnon branch into a two-magnon continuum.\cite{Thompson2017}. In Y-kapellasite, however, neither geometrical frustration nor anisotropic spin interactions seem to be present in the system \cite{Hering2021} excluding the magnon decay processes from the present case. Nevertheless, we note that a multi-center magnon based extraction of the SDOS with THz spectroscopy, as presented in this work, might be possible even in the presence of a decay mechanism.
\end{revision}

There are several issues concerning the magnetic properties of Y-kapellasite which should be addressed in future investigations. One particularly interesting \textcolor{red}{issue remaining} is the persistent spin dynamics in the magnetically ordered state below $T_N=2.2$~K \cite{Barthelemy19} which might be related to the successive spin freezing observed in the susceptibility, Figure~\ref{Fig1}c. For optical measurements, such a low temperature (mK range) is still challenging. Hence, complementary experimental techniques are required to scrutinize this exotic behavior. In addition, an investigation of the microscopic origin of the short-range magnetic correlations much above $T_N$ could be a focus of further studies.

\clearpage
\section{\textcolor{red}{Conclusion}}
\textcolor{red}{In conclusion, in our proof-of-principle experiment, we establish THz time-domain spectroscopy as a method} to directly probe the spin density of states expanding its capability from $\bm{q}\approx0$ to $\Delta \bm{q}\approx0$ excitations over the entire Brillouin zone: THz magnetometry. The \textcolor{red}{three-center magnon} absorption in the exotic magnetic superstructure of Y-kapellasite is the key mechanism behind this observation. Driven by short-range magnetic interactions, the absorption persists well above the magnetic ordering temperature, i.e. we do observe paramagnons. \textcolor{red}{The multi-center magnon absorption allows easy access to the spin density of states for suitable magnets, in particular in systems with low symmetry.}

\clearpage
\section{Experimental Section}
\threesubsection{Crystal growth and characterization}
Crystals were grown via a horizontal external gradient growth method. This optimized synthesis leads to nearly perfect defect-free single crystals with large facets (kagome-planes between 2 and 3 mm and thickness of around 0.5-1 mm), see supplement, Figure~\scrystal. The results of X-ray diffraction is presented in Figure~\sxray. The ac and dc magnetic susceptibilities were measured as explained in the supplement.

\threesubsection{THz-TDS measurements}
THz time-domain spectroscopy (THz-TDS) measures the time-dependent electric field.\cite{Neu2018} Through a Fast Fourier Transformation (FFT), intensity and phase can be obtained. This allows us to directly calculate the optical response functions.\cite{Tanner2019,dressel2002} A typical time-trace of the transient electric field shows an oscillating behavior with a strong pulse at early times (main pulse). Due to their short lifetime, electronic transitions or phononic resonances are, in general, contained in the main pulse. However, phenomena with a longer lifetime, such as magnetic resonances, can extend to longer times (over several tens of ps), exceeding the main pulse.\textcolor{red}{\cite{Lu17,Grishunin18,Nakajima2010,Yamaguchi2010,Yamaguchi2013}}

THz-TDS measurements were carried out in transmission geometry on oriented single crystals ($\bm{E}_{THz}$ $\parallel (a,b)$) at several temperatures between $295$ and $1.6$~K with a helium bath cryostat. Magneto-optical THz measurements were performed in Faraday geometry ($\bm{E}_{THz}$ $\parallel (a,b)$, $\bm{H}$ $\parallel c$) with static magnetic field strengths up to $H=10$~T and temperatures down to $1.7$~K. The absorption coefficient $\alpha$ was calculated from the transmittance {\it Tr} by $\alpha =-\mathrm{ln}\{\textit{Tr}\}/d$, where $d$ is the sample thickness. The integrated absorption coefficient $\textit{IA}=\int\alpha \,\mathrm{d}\omega$ resembles the optical spectral weight and provides a quantitative access to the spectral features (see supplement for further information). The frequency-dependent magnetic susceptibility $\chi^{''}_m=\mathrm{Im}\{\tilde{\chi}_m(\omega)\}$ was calculated by referencing to the high-temperature dielectric response,\cite{Kozuki2011,Laurita2015,Zhang18,Chauhan2020} i.e. the $T=80$~K spectrum, see supplement for further information. This quantity encodes the SDOS, as previously shown for the $\bm{q}\approx0$ case.\cite{Zhang18}

\threesubsection{ESR measurements}
\textcolor{red}{Temperature-dependent electron spin resonance (ESR)} measurements in the X-band frequency were carried out. The in-plane response was determined with a microwave field $\bm{h}\parallel c$ and external magnetic field $\bm{H}\parallel (a,b)$. More details are discussed in the supplement.

\threesubsection{NMR characterization}\textcolor{red}{Nuclear magnetic resonance ($^1$H-NMR)} experiments were performed on a $3\times 3 \times 1~{\rm mm}^3$ sized Y-kapellasite single crystal with magnetic field ($H = 0.98$~T) aligned parallel to the kagome layers ($\bm{H}\parallel (a,b)$). The spin-lattice relaxation rate $1/T_1$ was measured by non-selective excitation of the full line.  $1/T_1$ was determined through saturation-recovery using single-exponential fits. Temperature control in the range from 4 to 200~K was achieved using a $^4$He cryostat with a variable-temperature insert.

\threesubsection{High-field magnetization}
The high-field magnetization was determined between $T=0.5$ and $30$~K in pulsed magnetic fields up to $55$~T for in-plane and out-of-plane orientations. Further information can be found in the supplement. Measurements were performed at the high field laboratory in Dresden, Germany (HLD-EMFL).

\threesubsection{DFT and LSWT calculations}
The phonon frequencies were calculated using a combination of the PHONOPY package~\cite{Togo2008,Togo2015} and Density Functional Theory \textcolor{red}{(DFT)} as implemented in the \textcolor{red}{Vienna Ab-Initio Simulation Package (VASP) code.}\cite{Kresse1993, Kresse1996a, Kresse1996b} The details of the calculations are in the supplemental material.
The magnon dispersion was determined by using linear spin-wave theory (LSWT) as implemented in SpinW 3.0.\cite{Toth_2015} The details are explained in the supplement.
\clearpage

\textbf{Supporting Information} \par 
Supporting Information is available from the Wiley Online Library or from the author.

\medskip
\textbf{Acknowledgements} \par
We thank Artem Pronin, Guratinder Kaur, and Reinhard K. Kremer for fruitful discussion, Björn Miksch and Lena N. Majer for help with the ESR measurements and Gabriele Untereiner for continuous \textcolor{red}{technical} support. The project was supported by the HLD-HZDR, member of the European Magnetic Field Laboratory (EMFL), and the Deutsche Forschungsgemeinschaft (DFG). A.R. and R.V. thank the DFG through TRR 288-422213477 (project A05). Y.L. acknowledges support by China Postdoctoral Science Foundation (Grant No. 2019M660249) and National Natural Science Foundation of China (Grant No. 12004296). M.M. acknowledges the Swiss National Science Foundation (Grant No. 206021\_139082). A.P. acknowledges support by the Alexander von Humboldt Foundation through the Feodor Lynen Fellowship. The work at University of California, Los Angeles, was supported by NSF Grant 2004553.\\
\medskip

\textbf{Author contributions} \par
These authors contributed equally: T.B. and S.R. \\ T.B., S.R. performed the spectroscopic measurements and analyzed the data. P.P., K.M.Z, C.K. grew the crystals. P.P., J.W., S.S., D.G., Y.S., M.M. performed the magnetic characterization. J.N. performed the structural characterization.  A.P., S.E.B did the NMR measurements. A.R. performed the LSDW analysis. Y.L. calculated the phonon dispersion. A.P. contributed to the infrared measurements. T.B., S.R., P.P. wrote the manuscript with input from all authors. P.P., T.B., S.R. initiated the project. M.D. and R.V. supervised the project.\\
\medskip
\textbf{Competing Interests} \par
The authors declare that they have no competing financial interests.\\
\medskip
\textbf{Correspondence} \par
Correspondence and requests for materials should be addressed to \\
M.D. (dressel@pi1.physik.uni-stuttgart.de),\\ P.P. (p.puphal@fkf.mpg.de) or R.V. (valenti@itp.uni-frankfurt.de).

\clearpage

\begin{thebibliography}{10}
\providecommand{\url}[1]{\texttt{#1}}
\providecommand{\urlprefix}{URL }

\bibitem{dressel2002}
M.~Dressel, G.~Gr{\"u}ner,
\newblock \href{https://doi.org/10.1017/CBO9780511606168}{\emph{Electrodynamics of Solids: Optical Properties of Electrons in
  Matter},}
\newblock Cambridge University Press, Cambridge, \textbf{2002}.

\bibitem{Tanner2019}
D.~B. Tanner,
\newblock \href{https://doi.org/10.1017/9781316672778}{\emph{Optical Effects in Solids},}
\newblock Cambridge University Press, Cambridge, \textbf{2019}.

\bibitem{Fishman2018}
R.~S. Fishman, J.~A. Fernandez-Baca, T.~R{\~{o}}{\~{o}}m,
\newblock \href{https://doi.org/10.1088/978-1-64327-114-9}{\emph{{Spin-Wave Theory and its Applications to Neutron Scattering
  and THz Spectroscopy}},}
\newblock Morgan {\&} Claypool Publishers, San Rafael, \textbf{2018}.

\bibitem{Wulferding2019}
D.~Wulferding, Y.~Choi, W.~Lee, K.-Y. Choi,
\newblock \href{https://doi.org/10.1088/1361-648X/ab45c4}{\emph{J. Condens. Matter Phys.} \textbf{2019}, \emph{32}, 4 043001.}

\bibitem{Zhang18}
X.~Zhang, F.~Mahmood, M.~Daum, Z.~Dun, J.~A.~M. Paddison, N.~J. Laurita,
  T.~Hong, H.~Zhou, N.~P. Armitage, M.~Mourigal,
\newblock \href{https://doi.org/10.1103/PhysRevX.8.031001}{\emph{Phys. Rev. X} \textbf{2018}, \emph{8} 031001.}

\bibitem{Little17}
A.~Little, L.~Wu, P.~Lampen-Kelley, A.~Banerjee, S.~Patankar, D.~Rees, C.~A.
  Bridges, J.-Q. Yan, D.~Mandrus, S.~E. Nagler, J.~Orenstein,
\newblock \href{https://doi.org/10.1103/PhysRevLett.119.227201}{\emph{Phys. Rev. Lett.} \textbf{2017}, \emph{119} 227201.}

\bibitem{Chauhan2020}
P.~Chauhan, F.~Mahmood, H.~J. Changlani, S.~M. Koohpayeh, N.~P. Armitage,
\newblock \href{https://doi.org/10.1103/PhysRevLett.124.037203}{\emph{Phys. Rev. Lett.} \textbf{2020}, \emph{124}, 037203.}

\bibitem{Laurita2015}
N.~J. Laurita, J.~Deisenhofer, L.~Pan, C.~M. Morris, M.~Schmidt, M.~Johnsson,
  V.~Tsurkan, A.~Loidl, N.~P. Armitage,
\newblock \href{https://doi.org/10.1103/PhysRevLett.114.207201}{\emph{Phys. Rev. Lett.} \textbf{2015}, \emph{114} 207201.}

\bibitem{Nguyen2017}
T.~M.~H. Nguyen, L.~J. Sandilands, C.~H. Sohn, C.~H. Kim, A.~L. Wysocki, I.-S.
  Yang, S.~J. Moon, J.-H. Ko, J.~Yamaura, Z.~Hiroi, T.~W. Noh,
\newblock \href{https://doi.org/10.1038/s41467-017-00228-w}{\emph{Nat. Commun.} \textbf{2017}, \emph{8}, 1 251.}

\bibitem{Wu2018}
L.~Wu, A.~Little, E.~E. Aldape, D.~Rees, E.~Thewalt, P.~Lampen-Kelley,
  A.~Banerjee, C.~A. Bridges, J.-Q. Yan, D.~Boone, S.~Patankar,
  D.~Goldhaber-Gordon, D.~Mandrus, S.~E. Nagler, E.~Altman, J.~Orenstein,
\newblock \href{https://doi.org/10.1103/PhysRevB.98.094425}{\emph{Phys. Rev. B} \textbf{2018}, \emph{98} 094425.}

\bibitem{Hortensius2021}
J.~R. Hortensius, D.~Afanasiev, M.~Matthiesen, R.~Leenders, R.~Citro, A.~V. Kimel, R.~V. Mikhaylovskiy, B.~A. Ivanov, A.~D. Caviglia,
\newblock \href{https://doi.org/10.1038/s41567-021-01290-4}{\emph{Nat. Phys.} \textbf{2021}, \emph{17}, 9 1001.}

\bibitem{Walowski16}
J.~Walowski, M.~Münzenberg,
\newblock \href{https://doi.org/10.1063/1.4958846}{\emph{J. Appl. Phys.} \textbf{2016}, \emph{120}, 14 140901.}

\bibitem{Eschenlohr17}
A.~Eschenlohr, U.~Bovensiepen,
\newblock \href{https://doi.org/10.1088/1361-648X/aa9e69}{\emph{J. Condens. Matter Phys.} \textbf{2017}, \emph{30}, 3 030301.}

\bibitem{Nemec2018}
P.~N{\v{e}}mec, M.~Fiebig, T.~Kampfrath, A.~V. Kimel,
\newblock \href{https://doi.org/10.1038/s41567-018-0051-x}{\emph{Nat. Phys.} \textbf{2018}, \emph{14}, 3 229.}

\bibitem{Kampfrath13}
T.~Kampfrath, K.~Tanaka, K.~A. Nelson,
\newblock \href{https://doi.org/10.1038/nphoton.2013.184}{\emph{Nat. Photonics} \textbf{2013}, \emph{7}, 9 680.}

\bibitem{Kimel05}
A.~V. Kimel, A.~Kirilyuk, P.~A. Usachev, R.~V. Pisarev, A.~M. Balbashov,
  Th.~Rasing,
\newblock \href{https://doi.org/10.1038/nature03564}{\emph{Nature} \textbf{2005}, \emph{435}, 7042 655.}

\bibitem{Kampfrath11}
T.~Kampfrath, A.~Sell, G.~Klatt, A.~Pashkin, S.~M{\"a}hrlein, T.~Dekorsy,
  M.~Wolf, M.~Fiebig, A.~Leitenstorfer, R.~Huber,
\newblock \href{https://doi.org/10.1038/nphoton.2010.259}{\emph{Nat. Photonics} \textbf{2011}, \emph{5}, 1 31.}

\bibitem{Vicario13}
C.~Vicario, C.~Ruchert, F.~Ardana-Lamas, P.~M. Derlet, B.~Tudu, J.~Luning,
  C.~P. Hauri,
\newblock \href{https://doi.org/10.1038/nphoton.2013.209}{\emph{Nat. Photonics} \textbf{2013}, \emph{7}, 9 720.}

\bibitem{Hansteen2005}
F.~Hansteen, A.~Kimel, A.~Kirilyuk, T.~Rasing,
\newblock \href{https://doi.org/10.1103/PhysRevLett.95.047402}{\emph{Phys. Rev. Lett.} \textbf{2005}, \emph{95} 047402.}

\bibitem{Nakajima2010}
M.~Nakajima, A.~Namai, S.~Ohkoshi, T.~Suemoto,
\newblock \href{https://doi.org/10.1364/OE.18.018260}{\emph{Opt. Express} \textbf{2010}, \emph{18}, 17 18260.}

\bibitem{Yamaguchi2010}
K.~Yamaguchi, M.~Nakajima, T.~Suemoto,
\newblock \href{https://doi.org/10.1103/PhysRevLett.105.237201}{\emph{Phys. Rev. Lett.} \textbf{2010}, \emph{105} 237201.}

\bibitem{Yamaguchi2013}
K.~Yamaguchi, T.~Kurihara, Y.~Minami, M.~Nakajima, T.~Suemoto,
\newblock \href{https://doi.org/10.1103/PhysRevLett.110.137204}{\emph{Phys. Rev. Lett.} \textbf{2013}, \emph{110} 137204.}

\bibitem{Zhang2020}
W.~Zhang, P.~Maldonado, Z.~Jin, T.~S. Seifert, J.~Arabski, G.~Schmerber,
  E.~Beaurepaire, M.~Bonn, T.~Kampfrath, P.~M. Oppeneer, D.~Turchinovich,
\newblock \href{https://doi.org/10.1038/s41467-020-17935-6}{\emph{Nat. Commun.} \textbf{2020}, \emph{11}, 1 4247.}

\bibitem{Neu2018}
J.~Neu, C.~A. Schmuttenmaer,
\newblock \href{https://doi.org/10.1063/1.5047659}{\emph{J. Appl. Phys.} \textbf{2018}, \emph{124}, 23 231101.}

\bibitem{Wright1969}
G.~B. Wright, editor,
\newblock \href{https://doi.org/10.1007/978-3-642-87357-7}{\emph{{Light Scattering Spectra of Solids}},}
\newblock Springer Berlin Heidelberg, Berlin, Heidelberg, \textbf{1969}.

\bibitem{Fleury1968}
P.~A. Fleury, R.~Loudon,
\newblock \href{https://doi.org/10.1103/PhysRev.166.514}{\emph{Phys. Rev.} \textbf{1968}, \emph{166} 514.}

\bibitem{Loudon1968}
R.~Loudon,
\newblock \href{https://doi.org/10.1080/00018736800101296}{\emph{Adv. Phys.} \textbf{1968}, \emph{17}, 66 243.}

\bibitem{Rezende2019}
S.~M. Rezende, A.~Azevedo, R.~L. Rodr{\'{i}}guez-Su{\'{a}}rez,
\newblock \href{https://doi.org/10.1063/1.5109132}{\emph{J. Appl. Phys.} \textbf{2019}, \emph{126}, 15 151101.}

\bibitem{Grishunin18}
K.~Grishunin, T.~Huisman, G.~Li, E.~Mishina, T.~Rasing, A.~V. Kimel, K.~Zhang,
  Z.~Jin, S.~Cao, W.~Ren, G.-H. Ma, R.~V. Mikhaylovskiy,
\newblock \href{https://doi.org/10.1021/acsphotonics.7b01402}{\emph{ACS Photonics} \textbf{2018}, \emph{5}, 4 1375.}

\bibitem{Lu17}
J.~Lu, X.~Li, H.~Y. Hwang, B.~K. Ofori-Okai, T.~Kurihara, T.~Suemoto, K.~A.
  Nelson,
\newblock \href{https://doi.org/10.1103/PhysRevLett.118.207204}{\emph{Phys. Rev. Lett.} \textbf{2017}, \emph{118} 207204.}

\bibitem{Thorpe1970}
M.~F. Thorpe,
\newblock \href{https://doi.org/10.1063/1.1659005}{\emph{J. Appl. Phys.} \textbf{1970}, \emph{41}, 3 892.}

\bibitem{Azuma05}
S.~Azuma, M.~Sato, Y.~Fujimaki, S.~Uchida, Y.~Tanabe, E.~Hanamura,
\newblock \href{https://doi.org/10.1103/PhysRevB.71.014429}{\emph{Phys. Rev. B} \textbf{2005}, \emph{71} 014429.}

\bibitem{LeTacon2011}
M.~Le~Tacon, G.~Ghiringhelli, J.~Chaloupka, M.~Moretti Sala, V.~Hinkov, M.~W.
  Haverkort, M.~Minola, M.~Bakr, K.~J. Zhou, S.~Blanco-Canosa, C.~Monney, Y.~T.
  Song, G.~L. Sun, C.~T. Lin, G.~M. De~Luca, M.~Salluzzo, G.~Khaliullin,
  T.~Schmitt, L.~Braicovich, B.~Keimer,
\newblock \href{https://doi.org/10.1038/nphys2041}{\emph{Nat. Phys.} \textbf{2011}, \emph{7}, 9 725.}

\bibitem{Qin2017}
H.~J. Qin, Kh.~Zakeri, A.~Ernst, J.~Kirschner,
\newblock \href{https://doi.org/10.1103/PhysRevLett.118.127203}{\emph{Phys. Rev. Lett.} \textbf{2017}, \emph{118} 127203.}

\bibitem{Kadlec11}
C.~Kadlec, V.~Goian, K.~Z. Rushchanskii, P.~Ku\ifmmode~\check{z}\else
  \v{z}\fi{}el, M.~Le\ifmmode \check{z}\else
  \v{z}\fi{}ai\ifmmode~\acute{c}\else \'{c}\fi{}, K.~Kohn, R.~V. Pisarev,
  S.~Kamba,
\newblock \href{https://doi.org/10.1103/PhysRevB.84.174120}{\emph{Phys. Rev. B} \textbf{2011}, \emph{84} 174120.}

\bibitem{Balents2010}
L.~Balents,
\newblock \href{https://doi.org/10.1038/nature08917}{\emph{Nature} \textbf{2010}, \emph{464}, 7286 199.}

\bibitem{Mendels16}
P.~Mendels, F.~Bert,
\newblock \href{https://doi.org/10.1016/j.crhy.2015.12.001}{\emph{C. R. Phys.} \textbf{2016}, \emph{17}, 3 455.}

\bibitem{Kohama19}
Y.~Kohama, H.~Ishikawa, A.~Matsuo, K.~Kindo, N.~Shannon, Z.~Hiroi,
\newblock \href{https://doi.org/10.1073/pnas.1821969116}{\emph{Proc. Natl. Acad. Sci. U.S.A.} \textbf{2019}, \emph{116}, 22 10686.}

\bibitem{Norman16}
M.~R. Norman,
\newblock \href{https://doi.org/10.1103/RevModPhys.88.041002}{\emph{Rev. Mod. Phys.} \textbf{2016}, \emph{88} 041002.}

\bibitem{Knolle2019}
J.~Knolle, R.~Moessner,
\newblock \href{https://doi.org/10.1146/annurev-conmatphys-031218-013401}{\emph{Annu. Rev. Condens. Matter Phys.} \textbf{2019}, \emph{10}, 1 451.}

\bibitem{Puphal17}
P.~Puphal, M.~Bolte, D.~Sheptyakov, A.~Pustogow, K.~Kliemt, M.~Dressel,
  M.~Baenitz, C.~Krellner,
\newblock \href{https://doi.org/10.1039/C6TC05110C}{\emph{J. Mater. Chem. C} \textbf{2017}, \emph{5} 2629.}

\bibitem{Hering2021}
M.~Hering, F.~Ferrari, A.~Razpopov, I.~I. Mazin, R.~Valent{\'i}, H.~O. Jeschke,
  J.~Reuther,
\newblock \href{https://doi.org/10.1038/s41524-021-00689-0}{\emph{Npj Comput. Mater.} \textbf{2022}, \emph{8}, 1 10.}

\bibitem{Barthelemy19}
Q.~Barth\'elemy, P.~Puphal, K.~M. Zoch, C.~Krellner, H.~Luetkens, C.~Baines,
  D.~Sheptyakov, E.~Kermarrec, P.~Mendels, F.~Bert,
\newblock \href{https://doi.org/10.1103/PhysRevMaterials.3.074401}{\emph{Phys. Rev. Mater.} \textbf{2019}, \emph{3} 074401.}

\bibitem{Pustogow17}
A.~Pustogow, Y.~Li, I.~Voloshenko, P.~Puphal, C.~Krellner, I.~I. Mazin,
  M.~Dressel, R.~Valent\'{\i},
\newblock \href{https://doi.org/10.1103/PhysRevB.96.241114}{\emph{Phys. Rev. B} \textbf{2017}, \emph{96} 241114.}

\bibitem{Watanabe2016}
D.~Watanabe, K.~Sugii, M.~Shimozawa, Y.~Suzuki, T.~Yajima, H.~Ishikawa,
  Z.~Hiroi, T.~Shibauchi, Y.~Matsuda, M.~Yamashita,
\newblock \href{https://doi.org/10.1073/pnas.1524076113}{\emph{Proc. Natl. Acad. Sci. U.S.A.} \textbf{2016}, \emph{113}, 31
  8653.}

\bibitem{Ishikawa2019}
H.~Ishikawa, D.~Nishio-Hamane, A.~Miyake, M.~Tokunaga, A.~Matsuo, K.~Kindo,
  Z.~Hiroi,
\newblock \href{https://doi.org/10.1103/PhysRevMaterials.3.064414}{\emph{Phys. Rev. Mater.} \textbf{2019}, \emph{3} 064414.}

\bibitem{Boldrin2018}
D.~Boldrin, B.~F\aa{}k, E.~Can\'evet, J.~Ollivier, H.~C. Walker, P.~Manuel,
  D.~D. Khalyavin, A.~S. Wills,
\newblock \href{https://doi.org/10.1103/PhysRevLett.121.107203}{\emph{Phys. Rev. Lett.} \textbf{2018}, \emph{121} 107203.}

\bibitem{Pilon13}
D.~V. Pilon, C.~H. Lui, T.-H. Han, D.~Shrekenhamer, A.~J. Frenzel, W.~J.
  Padilla, Y.~S. Lee, N.~Gedik,
\newblock \href{https://doi.org/10.1103/PhysRevLett.111.127401}{\emph{Phys. Rev. Lett.} \textbf{2013}, \emph{111} 127401.}

\bibitem{Potter13}
A.~C. Potter, T.~Senthil, P.~A. Lee,
\newblock \href{https://doi.org/10.1103/PhysRevB.87.245106}{\emph{Phys. Rev. B} \textbf{2013}, \emph{87} 245106.}

\bibitem{Windt01}
M.~Windt, M.~Gr\"uninger, T.~Nunner, C.~Knetter, K.~P. Schmidt, G.~S. Uhrig,
  T.~Kopp, A.~Freimuth, U.~Ammerahl, B.~B\"uchner, A.~Revcolevschi,
\newblock \href{https://doi.org/10.1103/PhysRevLett.87.127002}{\emph{Phys. Rev. Lett.} \textbf{2001}, \emph{87} 127002.}

\bibitem{grueninger20}
M.~Gr\"uninger, D.~van~der Marel, A.~Damascelli, A.~Erb, T.~Nunner, T.~Kopp,
\newblock \href{https://doi.org/10.1103/PhysRevB.62.12422}{\emph{Phys. Rev. B} \textbf{2000}, \emph{62} 12422.}

\bibitem{Lorenzana1995}
J.~Lorenzana, G.~A. Sawatzky,
\newblock \href{https://doi.org/10.1103/PhysRevLett.74.1867}{\emph{Phys. Rev. Lett.} \textbf{1995}, \emph{74} 1867.}

\bibitem{Perkins1993}
J.~D. Perkins, J.~M. Graybeal, M.~A. Kastner, R.~J. Birgeneau, J.~P. Falck,
  M.~Greven,
\newblock \href{https://doi.org/10.1103/PhysRevLett.71.1621}{\emph{Phys. Rev. Lett.} \textbf{1993}, \emph{71} 1621.}

\bibitem{Halley1965}
J.~W. Halley, I.~Silvera,
\newblock \href{https://doi.org/10.1103/PhysRevLett.15.654}{\emph{Phys. Rev. Lett.} \textbf{1965}, \emph{15} 654.}

\bibitem{Tanabe1965}
Y.~Tanabe, T.~Moriya, S.~Sugano,
\newblock \href{https://doi.org/10.1103/PhysRevLett.15.1023}{\emph{Phys. Rev. Lett.} \textbf{1965}, \emph{15} 1023.}

\bibitem{Allen1966}
S.~J. Allen, R.~Loudon, P.~L. Richards,
\newblock \href{https://doi.org/10.1103/PhysRevLett.16.463}{\emph{Phys. Rev. Lett.} \textbf{1966}, \emph{16} 463.}

\bibitem{Thorpe1969}
M.~F. Thorpe, R.~J. Elliott,
\newblock In G.~B. Wright, editor, \href{https://doi.org/10.1007/978-3-642-87357-7_21}{\emph{Light Scattering Spectra of Solids},
  chapter C-2, 199--206.} Springer Berlin Heidelberg, Berlin, Heidelberg,
  \textbf{1969}.

\bibitem{Dexter1962}
D.~L. Dexter,
\newblock \href{https://doi.org/10.1103/PhysRev.126.1962}{\emph{Phys. Rev.} \textbf{1962}, \emph{126} 1962.}

\bibitem{Halley1967}
J.~W. Halley,
\newblock \href{https://doi.org/10.1103/PhysRev.154.458}{\emph{Phys. Rev.} \textbf{1967}, \emph{154} 458.}

\bibitem{Lohr1972}
L.~{L. Lohr},
\newblock \href{https://doi.org/10.1016/S0010-8545(00)80030-2}{\emph{Coord. Chem. Rev.} \textbf{1972}, \emph{8}, 3 241.}

\bibitem{Tanabe2005}
Y.~Tanabe, E.~Hanamura,
\newblock \href{https://doi.org/10.1143/JPSJ.74.670}{\emph{J. Phys. Soc. Japan} \textbf{2005}, \emph{74}, 2 670.}

\bibitem{Zhitomirsky13}
M.~E. Zhitomirsky, A.~L. Chernyshev,
\newblock \href{https://doi.org/10.1103/RevModPhys.85.219}{\emph{Rev. Mod. Phys.} \textbf{2013}, \emph{85} 219.}

\bibitem{Thompson2017}
J.~D. Thompson, P.~A. McClarty, D.~Prabhakaran, I.~Cabrera, T.~Guidi,
  R.~Coldea,
\newblock \href{https://doi.org/10.1103/PhysRevLett.119.057203}{\emph{Phys. Rev. Lett.} \textbf{2017}, \emph{119} 057203.}

\bibitem{Hong2017}
T.~Hong, Y.~Qiu, M.~Matsumoto, D.~A. Tennant, K.~Coester, K.~P. Schmidt, F.~F.
  Awwadi, M.~M. Turnbull, H.~Agrawal, A.~L. Chernyshev,
\newblock \href{https://doi.org/10.1038/ncomms15148}{\emph{Nat. Commun.} \textbf{2017}, \emph{8}, 1 15148.}

\bibitem{Wulferding2020}
D.~Wulferding, Y.~Choi, S.-H. Do, C.~H. Lee, P.~Lemmens, C.~Faugeras,
  Y.~Gallais, K.-Y. Choi,
\newblock \href{https://doi.org/10.1038/s41467-020-15370-1}{\emph{Nat. Commun.} \textbf{2020}, \emph{11}, 1 1603.}

\bibitem{Sahasrabudhe20}
A.~Sahasrabudhe, D.~A.~S. Kaib, S.~Reschke, R.~German, T.~C. Koethe, J.~Buhot,
  D.~Kamenskyi, C.~Hickey, P.~Becker, V.~Tsurkan, A.~Loidl, S.~H. Do, K.~Y.
  Choi, M.~Gr\"uninger, S.~M. Winter, Z.~Wang, R.~Valent\'{\i}, P.~H.~M. van
  Loosdrecht,
\newblock \href{https://doi.org/10.1103/PhysRevB.101.140410}{\emph{Phys. Rev. B} \textbf{2020}, \emph{101} 140410.}

\bibitem{Pan2014}
L.~Pan, S.~K. Kim, A.~Ghosh, C.~M. Morris, K.~A. Ross, E.~Kermarrec, B.~D.
  Gaulin, S.~M. Koohpayeh, O.~Tchernyshyov, N.~P. Armitage,
\newblock \href{https://doi.org/10.1038/ncomms5970}{\emph{Nat. Commun.} \textbf{2014}, \emph{5}, 1 4970.}

\bibitem{Winter2017}
S.~M. Winter, K.~Riedl, P.~A. Maksimov, A.~L. Chernyshev, A.~Honecker,
  R.~Valent{\'i},
\newblock \href{https://doi.org/10.1038/s41467-017-01177-0}{\emph{Nat. Commun.} \textbf{2017}, \emph{8}, 1 1152.}


\bibitem{Wang17}
Z.~Wang, S.~Reschke, D.~H\"uvonen, S.-H. Do, K.-Y. Choi, M.~Gensch, U.~Nagel,
  T.~R{\~{o}}{\~{o}}m, A.~Loidl,
\newblock \href{https://doi.org/10.1103/PhysRevLett.119.227202}{\emph{Phys. Rev. Lett.} \textbf{2017}, \emph{119} 227202.}

\bibitem{Kozuki2011}
K.~Kozuki, T.~Nagashima, M.~Hangyo,
\newblock \href{https://doi.org/10.1364/OE.19.024950}{\emph{Opt. Express} \textbf{2011}, \emph{19}, 25 24950.}

\bibitem{Togo2008}
A.~Togo, F.~Oba, I.~Tanaka,
\newblock \href{https://doi.org/10.1103/PhysRevB.78.134106}{\emph{Phys. Rev. B} \textbf{2008}, \emph{78} 134106.}

\bibitem{Togo2015}
A.~Togo, I.~Tanaka,
\newblock \href{https://doi.org/10.1016/j.scriptamat.2015.07.021}{\emph{Scr. Mater.} \textbf{2015}, \emph{108} 1.}

\bibitem{Kresse1993}
G.~Kresse, J.~Hafner,
\newblock \href{https://doi.org/10.1103/PhysRevB.47.558}{\emph{Phys. Rev. B} \textbf{1993}, \emph{47} 558.}

\bibitem{Kresse1996a}
G.~Kresse, J.~Furthm\"uller,
\newblock \href{https://doi.org/10.1103/PhysRevB.54.11169}{\emph{Phys. Rev. B} \textbf{1996}, \emph{54} 11169.}

\bibitem{Kresse1996b}
G.~Kresse, J.~Furthm\"uller,
\newblock \href{https://doi.org/10.1016/0927-0256(96)00008-0}{\emph{Comput. Mater. Sci.} \textbf{1996}, \emph{6}, 1 15.}

\bibitem{Toth_2015}
S.~Toth, B.~Lake,
\newblock \href{https://doi.org/10.1088/0953-8984/27/16/166002}{\emph{J. Condens. Matter Phys.} \textbf{2015}, \emph{27}, 16 166002.}

\end{thebibliography}

\clearpage
\lhead{}

\clearpage
\phantomsection\addcontentsline{toc}{section}{Supplement}
\includepdf[pages=-]{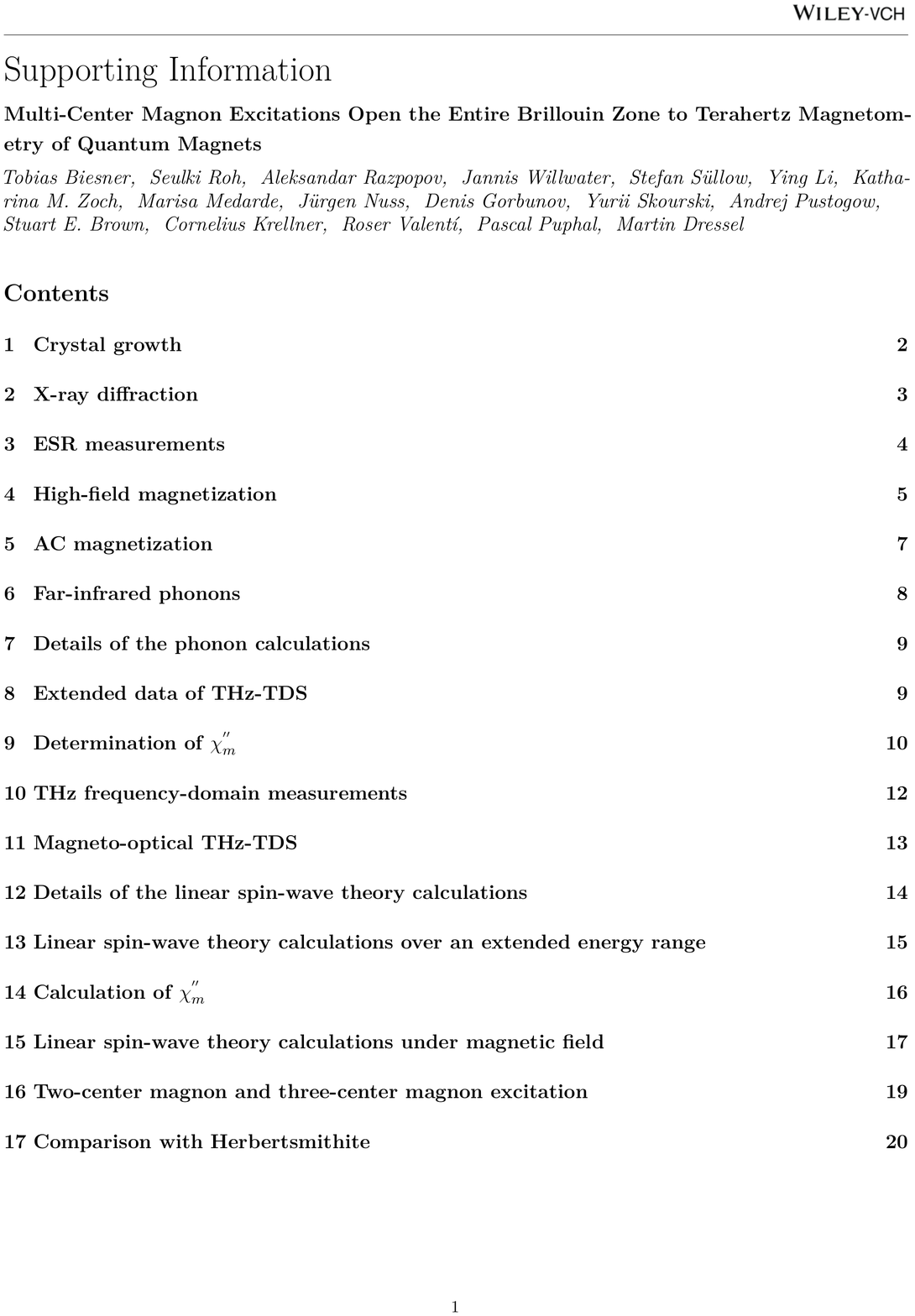}

\end{document}